\documentclass[twocolumn,english,nolinenumbers,aps,twocolumn]{revtex4-1}
\usepackage[T1]{fontenc}
\usepackage[utf8]{inputenc}
\usepackage{color}
\usepackage{babel}
\usepackage{amssymb}
\usepackage{graphicx}
\usepackage[unicode=true,pdfusetitle,
 bookmarks=true,bookmarksnumbered=false,bookmarksopen=false,
 breaklinks=false,pdfborder={0 0 0},backref=false,colorlinks=true]
 {hyperref}
\hypersetup{
 citecolor=blue}

\makeatletter
\usepackage{babel}

\makeatother

\begin{document}

\title{Regenerative memory in time-delayed neuromorphic photonic systems}

\author{B. Romeira$^{(1,2)}$, R. Avó$^{(2)}$, Jos\'{e} M. L. Figueiredo$^{(2)}$,
S. Barland$^{(3)}$ and J. Javaloyes$^{(4)}$}

\affiliation{$^{(1)}$ Centro de electrónica, Optoelectrónica y telecomunicações
(CEOT), Departmento de Física, Universidade do Algarve, 8005-139,
Faro, Portugal\\
$^{(2)}$ COBRA Research Institute, Eindhoven University of Technology,
P.O. Box 513, NL-5600 MB Eindhoven, Netherlands \\
 $^{(3)}$ Institut Non-Linéaire de Nice, Université de Nice Sophia
Antipolis, CNRS UMR 7335, 06560 Valbonne, France\\
$^{\left(4\right)}$Departament de Física, Universitat de les Illes
Baleares, C/ Valldemossa km 7.5, 07122 Mallorca, Spain}
\begin{abstract}
We investigate a regenerative memory based upon a time-delayed neuromorphic
photonic oscillator and discuss the link with temporal localized structures.
Our experimental implementation is based upon a optoelectronic system
composed of a nanoscale nonlinear resonant tunneling diode coupled
to a laser that we link to the paradigm of neuronal activity, the
FitzHugh-Nagumo model with delayed feedback.
\end{abstract}
\maketitle
Self-feedback connections in neurons are common in the nervous system
and are named autapses \cite{V-BR-72,HK-IJBC-04,F-NRN-09,B-NRN-09}.
These synapses between a neuron and a branch of its own axon are ubiquitous
and have been found in the neocortex and the hippocampus, to cite
but a few \cite{V-BR-72}. Yet their purpose has remained uncertain.
Recent reports suggest that autaptic transmission neurons are involved
in the long-lasting response to brief stimulation \cite{F-NRN-09},
and that this persistent activity has important implications in local
feedback regulation \cite{B-NRN-09}, and working memory. On the other
hand, recent progresses in multidisciplinary fields including semiconductor
physics, Photonics, computing and networking yielded the possibility
to emulate some elementary functions of the brain using neuromorphic
systems \cite{JCE-NL-10,ILB-FRNE-11,MAA-SCI-14,TNS-JLT-14}. The central
goal is to reproduce neuronal synapses by interconnecting thousands
of neuron-like elements \cite{KJL-NL-11}. While a lot of attention
has been dedicated to the network architecture of neuromorphic systems
\cite{MAA-SCI-14,TNS-JLT-14}, almost no attention has being paid
to the self-feedback autaptic connections providing self-localized
persistence of neuronal activity. 

In this work, we demonstrate temporal localized structures in a bio-inspired
time-delayed neuromorphic photonic system that exhibits long-lasting
responses and short transients to brief stimulation that can be explored
in regenerative memory and information storage. Because we use light
at telecommunications wavelengths for the regeneration of the spike-neuron
signals, we anticipate the development of fast neuron-inspired optical
storage interconnects and signal processing applications. We reduce
our physical experimental system to the prototypical FitzHugh-Nagumo
(FHN) model, a paradigm of neuronal response, complemented with a
self-feedback autaptic connection, see Fig.~1. This analysis bridges
our physical photonic system with the biological neuron and extend
our findings to other biological, physical and engineering systems. 

\begin{figure*}[t]
\begin{centering}
\includegraphics[width=0.66\columnwidth]{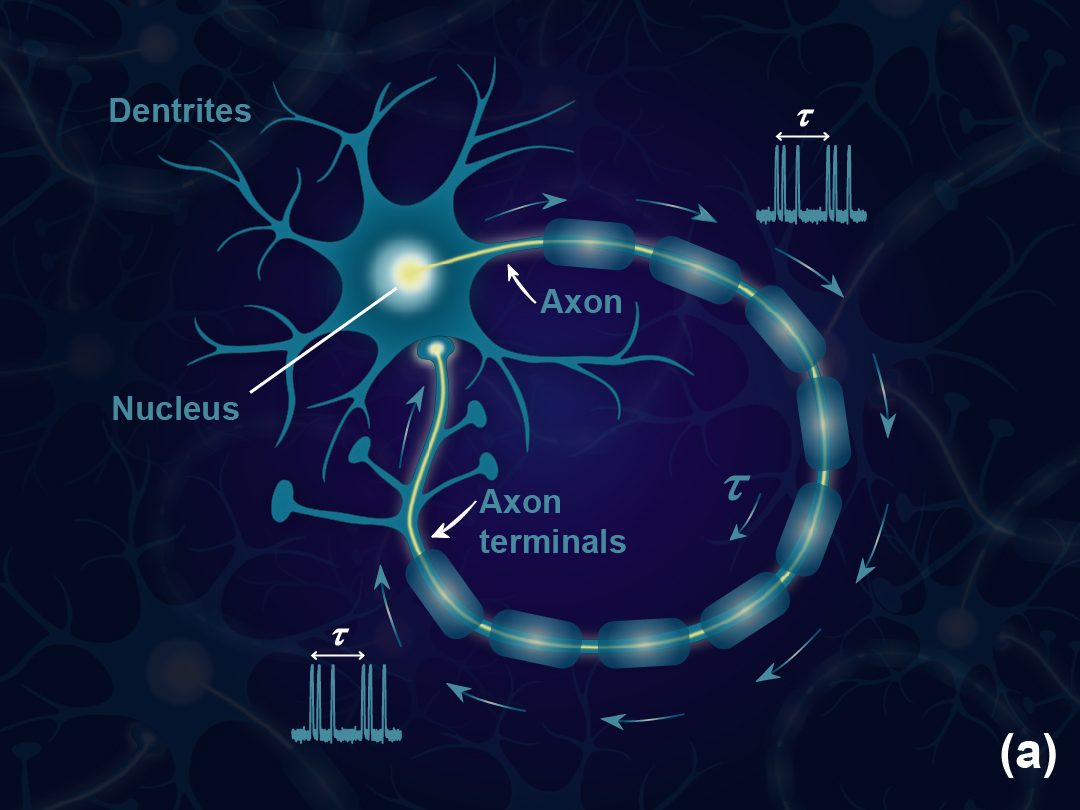}\enskip{}\includegraphics[width=0.66\columnwidth]{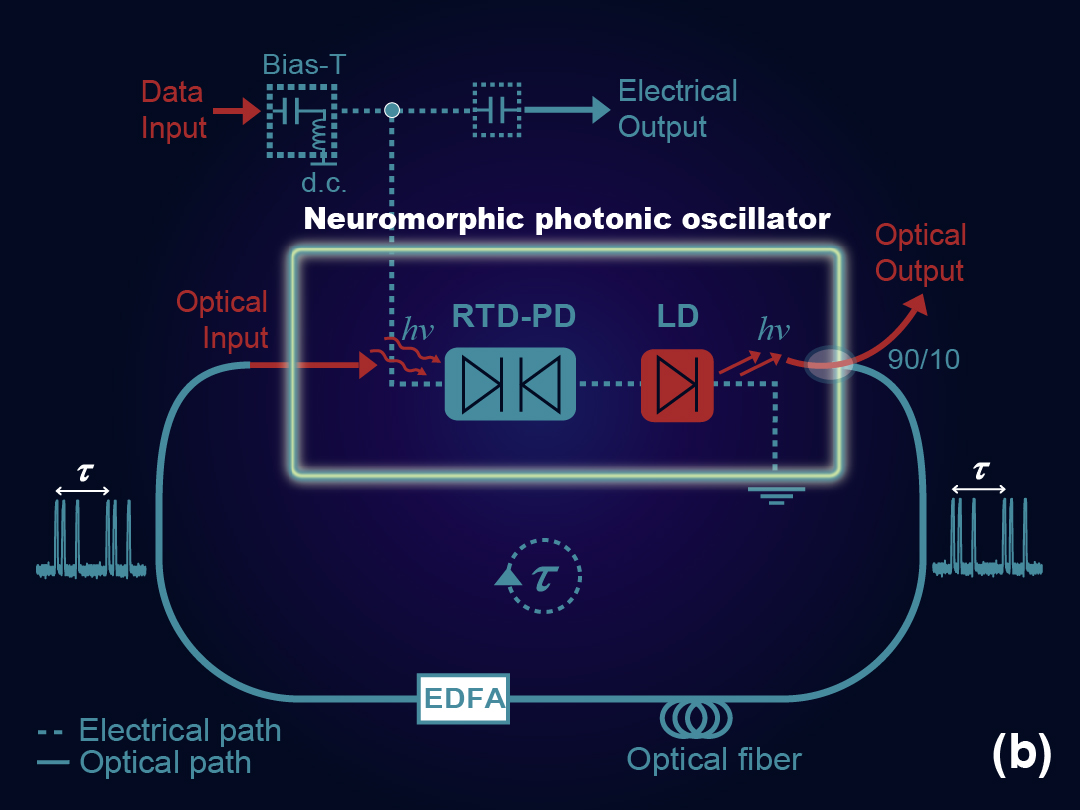}\enskip{}\includegraphics[width=0.66\columnwidth]{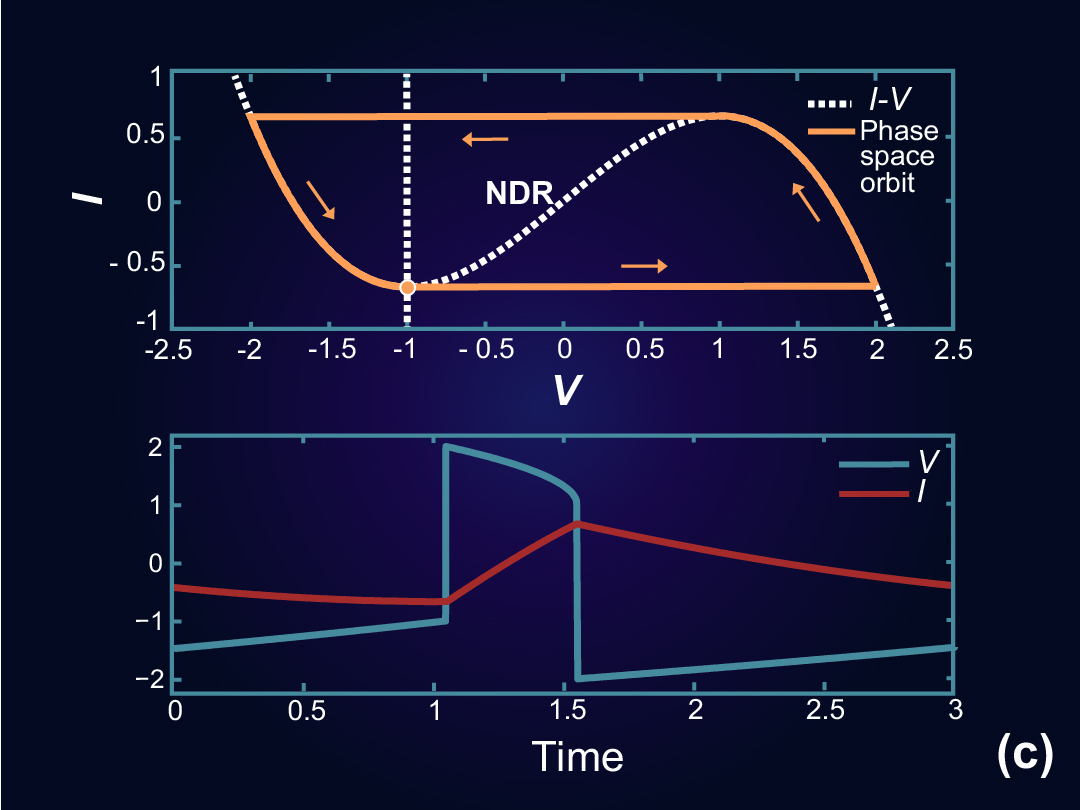}
\par\end{centering}

\centering{}\caption{(a) Diagram of a neuron with a self-feedback effect due to the presence
of an autapse. (b) Schematic of the equivalent time-delayed neuromorphic
photonic system. (c) nullclines and phase space trajectory of the
FHN model.\label{Fig1}}
\end{figure*}

The global behavior of large-scale neural assemblies \cite{E-RPP-98}
and how the brain activity produces higher cognition that can be emulated
in physical neuromorphic technologies leads in particular to the question
of memory and how information is stored. While, the complexity of
the brain is usually ascribed to the gigantic number of axons defining
the interconnections, the finite transit time of the electrical information
between individual units is also a source of complexity. A lot of
interested was devoted during the past decades to the effects of communication
time delays \cite{BT-PRE-03,S-PTA-09} and how they can influence
the dynamics. For instance, it was shown that dynamical systems mimicking
coupled neurons \cite{YMG-PRE-02,KBS-PRE-10,WEK-PRE-14} exhibit,
instead of a steady state, stable periodic pulsating regimes where
the two neurons may release energy in anti-phase, the period being
related to the delay value. However, the presence of time delays in
the dynamics may have a much deeper influence than to induce oscillations.
Even a \emph{single} delayed equation is akin to an infinite dimensional
dynamical system and important conceptual links between partial differential
equations (PDE)s and delayed differential equations (DDE)s exist \cite{GP-PRL-96}.
As such, one is lead to wonder if the rich dynamics found in spatially
extended systems, and in particular their ability to store information
\cite{CoulletLSinfo}, could also exists in the temporal output of
a \emph{single }neuron with a delayed coupling representing an autaptic
connection as seen in Fig.~\ref{Fig1}a).

One of the most interesting aspect of spatially extended out of equilibrium
nonlinear systems is their ability to generate complex spatial shapes,
e.g. Turing patterns \cite{turing52}, and Localized Structures (LS).
Because these LS coexist with the background homogeneous solution,
they can be individually addressed and used as bits of information.
Localized structures appear in a dissipative environment as attractors,
i.e. stable solutions towards which the system will evolve spontaneously
from a wide set of initial conditions \cite{NP-SelfOrg-77,DC-LNP-11}
making them intrinsically robust and allowing for complex interactions.
Localized states have been widely observed in nature in systems like
granular media \cite{sands}, gas discharges \cite{Astrov2001349},
semiconductor devices \cite{NAD-PSS-92}, reaction-diffusion systems
\cite{reactiondiffusion}, fluids \cite{PhysRevLett.52.1421}, convective
systems \cite{PhysRevA.35.2757} and optical cavities \cite{BTB-NAT-02,HPC-JSTQE-06,LCK-NAP-10,GJT-NC-15}.
Neuromorphic responses were achieved in lasers \cite{BKY-OL-11} and
nanolasers \cite{SBB-PRL-14} recently and, while in principle fundamentally
related to the existence of spatial degrees of freedom, Localized
States analogues have recently been observed in delayed dynamical
systems \cite{MGB-PRL-14,GJT-NC-15}.

Our time-delayed system consists of a nanoscale negative differential
conductance device based on a resonant tunneling photo-detector (RTD-PD)
diode set in an excitable regime and driving an integrated laser diode
source (LD). Although electronic neuron-like semiconductor microstructures
\cite{SNJ-JAP-11} have been proposed, they operate at rather low-speeds
(of the order of 20 kHz) and do not possess optical input/output.
The schematic diagram of our experimental neuromorphic photonic memory
is depicted in Fig.~\ref{Fig1}(b). The RTD-PD-LD provides a non-monotonic
Current-Voltage (I-V) curve with a region of negative differential
resistance, see Fig.~\ref{Fig1}(c), in which it behaves as an excitable
system. Indeed, as demonstrated in \cite{RJI-OE-13}, it provides
an all-or-none response characteristic of neurons. The RTD response
drives the laser diode which produces an optical pulse.

\begin{figure}
\centering{}\includegraphics[bb=35bp 0bp 420bp 320bp,clip,width=1\columnwidth]{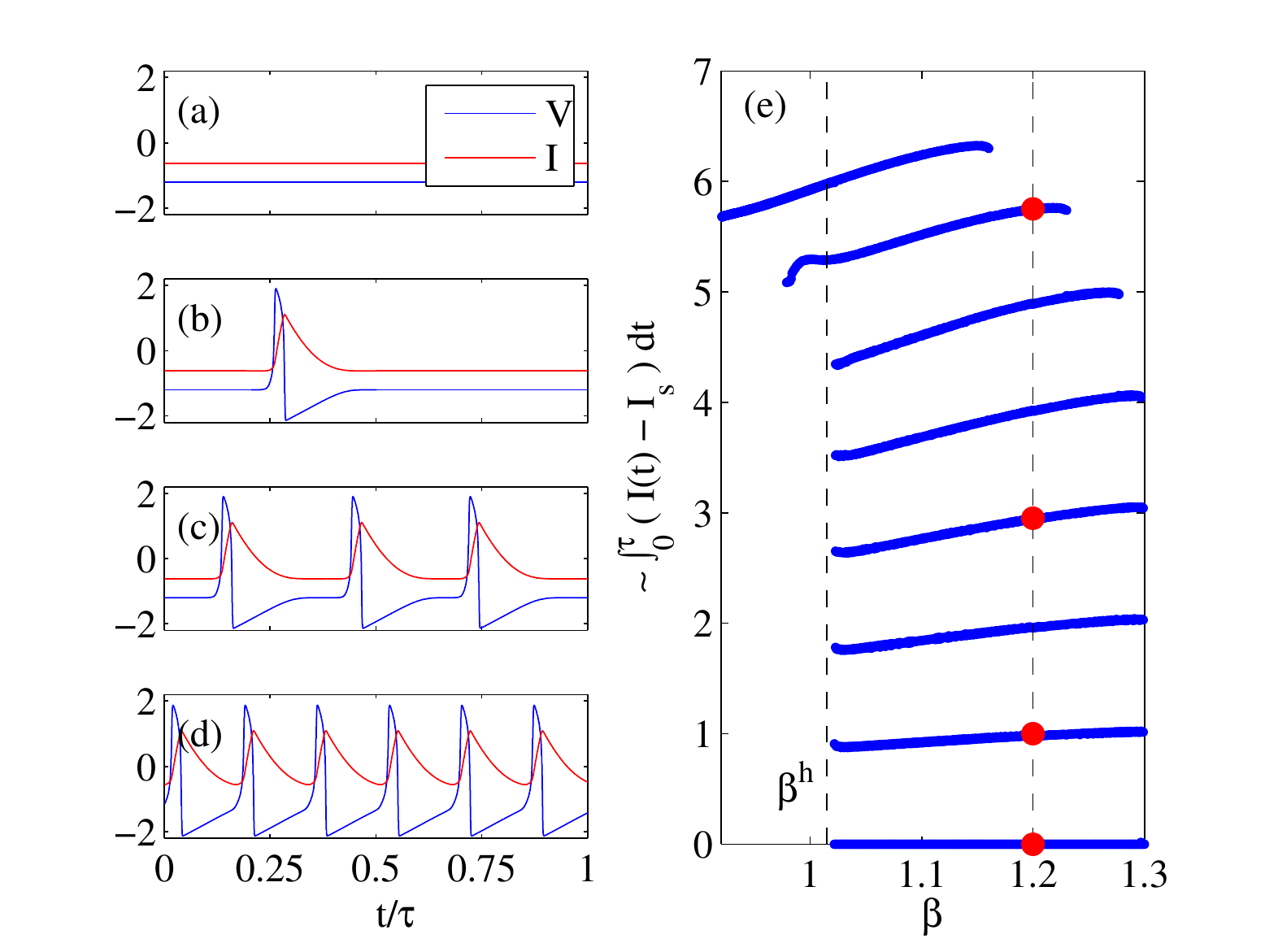}\caption{(Color online) Temporal time traces over a single period for various
states of the regenerative memory (a-d) and multi-stability diagram
of the coexisting solutions (e). In (e) we represent some norm of
the solutions as the integral over one period of the deviation of
the slow variable ($I-I_{s}$) yielding upward pulses with a zero
background. This integral is normalized to a value of $1$ when there
is a single pulse at $\beta=1.3$. All the localized solutions becomes
unstable close to $\beta^{h}\sim1.02$ where the background get destabilized
trough an Andronov-Hopf bifurcation. \label{Fig2} }
\end{figure}

Excitability is a concept originally coined to describe the capacity
of living organisms e.g. nerves \cite{HH-JOP-52,F-BMB-55,NAY-PIRE-62}
or neurons to respond strongly to a weak external stimulus that overcome
a well defined threshold. If the system is perturbed from its rest
state, it may relax back toward it steady state in two different ways.
If the perturbation remains below a certain threshold, the relaxation
is exponential. Above the threshold, the system has to perform a large
orbit that involves the whole phase space topology before relaxing
again toward the unique fixed point. Such a relaxation is visible
for instance in the lower panel of Fig.~\ref{Fig2}c). These two
widely different transitory regimes toward an unique attractor are
what defines the so-called excitability phenomenon. During its large
excursion in phase space, the system cannot respond to another perturbation
which defines incidentally the so-called lethargic time $T_{l}$ as
the temporal extend of the orbit. Well known in physiology, this refractory
period corresponds physically to the fact that a large amount of energy
is released during the excitable response and may be understood as
the time the system needs to recharge before being able to release
another response. In neurons, this period of time occurs during the
re-polarization and the hyperpolarization of the membrane potential. 

In order to operate our neuromorphic photonic system as a regenerative
memory, an optical delay line provides the temporal buffer memory
that will store the bits of information as light intensity pulses.
The excitable response of the RTD ensures the regeneration and the
rectification of the signal as a special nonlinear node along the
propagation loop. An erbium doped fiber amplifier (EDFA) is employed
in the loop to control the amount of feedback coupling and compensate
for the losses incurred by coupling and decoupling light from the
RTD-PD and LD chips (see methods A for details). The light pulse re-injection
into the RTD-PD triggers, after the round-trip in the fiber, a new
electrical response in the nano-optoelectronic system thereby repeating
the cycle with a period close to the optical pulse propagation time,
$\tau$, in the fiber. 

A precise modeling of the experimental situation can be achieved within
the framework of a dynamical model employing a Liénard equation \cite{L-RGE-28}
describing the RTD-PD oscillator \cite{RJF-JQE-13,RJI-OQEL-14} coupled
to the single mode rate equations modeling the laser intensity and
its population inversion (see method B). In the absence of feedback,
such an approach yields quantitative agreement with the experimental
results \cite{RJI-OE-13}. However, by assuming that the excitable
response is slower than the relaxation oscillation frequency of the
laser, one can adiabatically eliminate the laser intensity ($S$)
that becomes slaved to the current ($I$) of the RTD. By expanding
the nonlinear characteristic of the RTD at the center of the negative
differential resistance, denoting $V$ the deviation of the voltage
and assuming it is antisymmetric, one obtains exactly the FHN model,
making a complete link with our time-delayed neuromorphic photonic
system and the paradigm of excitability (see method B for more details).
The FHN model \cite{F-BMB-55,NAY-PIRE-62} represents a simplified
version of the theory developed by Hodgkin and Huxley \cite{HH-JOP-52,HH2-JOP-52}
to study how action potentials in neurons are initiated and propagated.
In order to allow for memory and complex dynamics, we introduce in
the FHN model a delayed perturbation. As such, the equations read
in their simplest form
\begin{eqnarray}
\dot{V} & = & V-\frac{V^{3}}{3}-I+\eta\left[I\left(t-\tau\right)-I\right],\label{eq:FHN1}\\
\dot{I} & = & \varepsilon\left(\beta+V\right).\label{eq:FHN2}
\end{eqnarray}

The stiffness parameters $\varepsilon$ denotes the ratio of the time
scale governing the slow $\left(I\right)$ and the fast $\left(V\right)$
variables while $\beta$ is the bias parameter. We choose $\beta>0$
without loss of generality. The amplitude of the delayed feedback
is denoted $\eta$. For the sake of convenience we use the so-called
form of non invasive feedback \cite{P-PLA-92}. As such, the steady
states of the FHN model are unchanged by the presence of feedback.
In correspondence with the experimental situation, we re-inject the
slow variable after a time delay $\tau$, into the dynamics of the
fast one as $I\left(t-\tau\right),$ yet very similar results were
obtained in other situations, e.g. re-injecting $V\left(t-\tau\right)$
instead. If not otherwise stated the parameters are $\varepsilon=0.05$,
$\eta=0.18$ and $\tau=500$. We included white Gaussian noise of
variable amplitude $\xi$ to model the stochastic processes occurring
in our experimental neuromorphic photonic oscillator.

We briefly recall the main properties of Eqs.~(\ref{eq:FHN1},\ref{eq:FHN2})
in the absence of delayed feedback. For values of the bias $\beta\gtrsim1$,
the unique steady state $\left(V_{s},I_{s}\right)=\left(-\beta,\beta^{3}/3-\beta\right)$
is stable like e.g. in Fig.~\ref{Fig2}a) yet the system is excitable.
For $\beta^{*}=1$ an Andronov-Hopf bifurcation occurs at frequency
$\omega^{*}=\sqrt{\varepsilon}$ yielding a weakly nonlinear oscillations
of the background solution which rapidly develops into a large amplitude
cycle for $\beta<\beta^{*}$ via the so-called canard phenomenon \cite{BCD-CM-81}.
In the latter case, the Eqs.~(\ref{eq:FHN1},\ref{eq:FHN2}) represents
a so-called relaxation oscillator \cite{keener2008,Meron19921}.

We are interested in the regimes in which the values of the time delay
is large as compared to the lethargic time $T_{l}$ which in our case
is $T_{l}\sim3\varepsilon^{-1}$, see \cite{Tyson1988327} for more
details. We demonstrate in Fig.~\ref{Fig2} that Eqs.~(\ref{eq:FHN1},\ref{eq:FHN2})
support in this case the storage of information. The presence of a
weak delayed perturbation in Eqs.~(\ref{eq:FHN1},\ref{eq:FHN2})
induces a multi-stability between an infinity of different temporal
patterns that repeat themselves identically with a period close to
the time delay $\tau$. For a broad range of values of the feedback
rate $\eta$, for which the re-injection of the delayed excitable
orbit after a time delay $\tau$ is sufficient to overcome the excitability
threshold, a new, perfectly formed, excitable orbit gets regenerated.
As such, the excitable orbit of the solitary FHN system, whose temporal
extend is $T_{l}$, becomes a binary unit of information \emph{embedded}
in a much longer periodic orbit of period $\tau$. Such bits of information
get perfectly regenerated after each period and since the mere condition
for \emph{perfect} regeneration is to overcome the excitability threshold,
one foresees the signal healing properties and the robustness of this
mechanism. We exemplify in Fig.~\ref{Fig2}a-d) various occurrences
of such periodic regimes whose period are close to $\tau$ and that
are composed of 0, 1, 3 and 6 bits of information as embedded excitable
responses within the time delay $\tau$. We stress in Figure~\ref{Fig2}e)
that all these regimes coexist between themselves for a wide range
of the bias parameter $\beta$. Because these isolated temporal patterns
are bistable with the uniform state and are also independent of the
boundary conditions, i.e. the time delay value, and are attractor
of the dynamics, they can be considered as the equivalent of Localized
Structures in time delayed systems.

\begin{figure}
\centering{}\includegraphics[bb=0bp 0bp 420bp 320bp,clip,width=1\columnwidth]{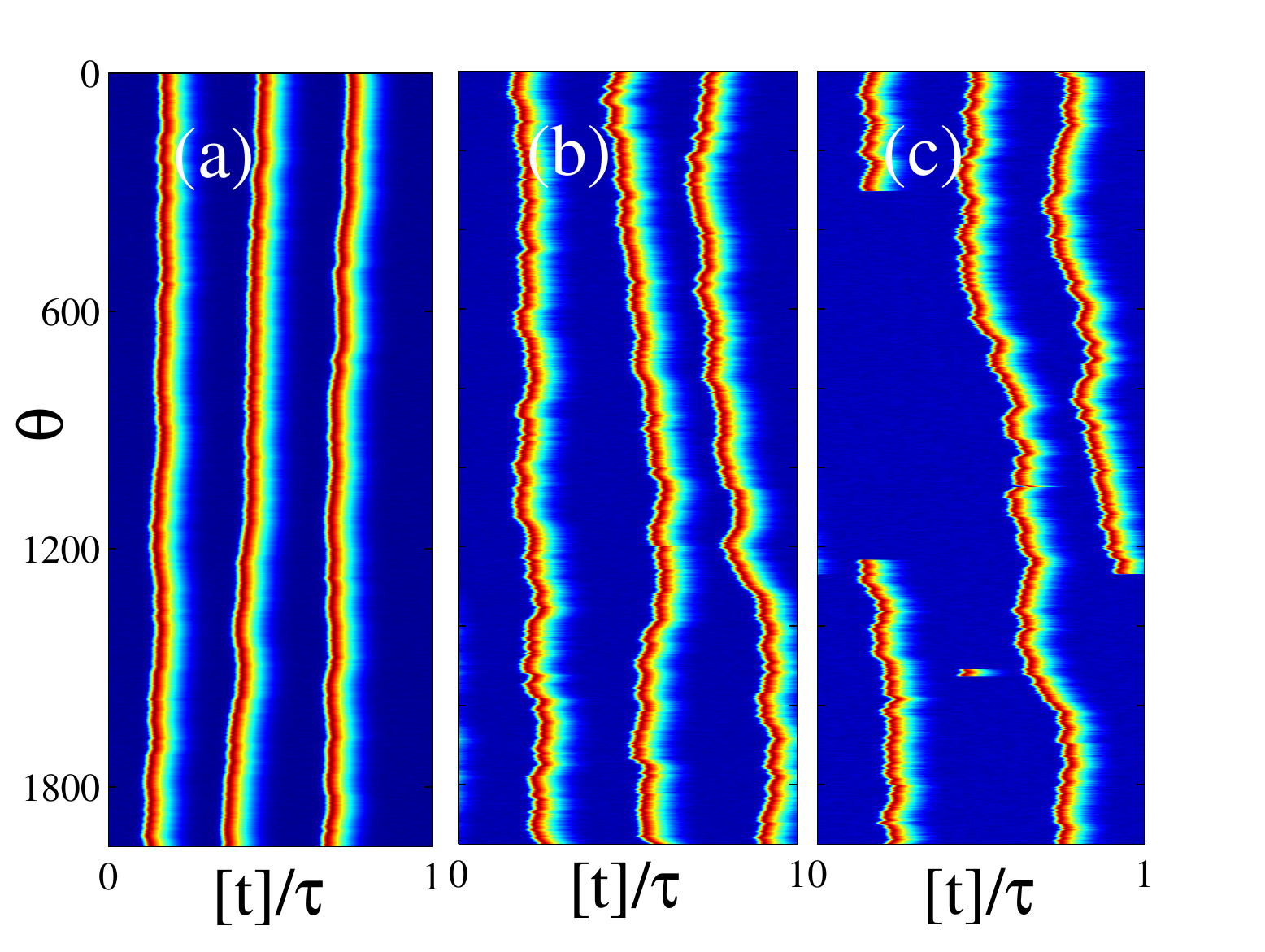}\caption{(Color online) The space-time plots represent the evolution of the
solution depicted in Fig.~\ref{Fig2}c) with three LS for an increasing
level of noise. The slow variable represented is $I$. In panel a)
$\xi=10^{-3}$ and the noise induced drift motion is barely visible
over more than $\theta=2000$ periods. In panel b) the uncorrelated
random walk is more visible since $\xi=5\times10^{-3}$ while in c)
the huge level of noise $\xi=8\times10^{-3}$ is capable of inducing
annihilation and nucleation of the LS. \label{Fig3} }
\end{figure}

We also depict in Fig.~\ref{Fig2}e the norm of the various solutions
as well the maximal number of bits of information that can be stored
for a given value of $\tau$. Interestingly, each branch of solution
corresponds to a well defined number of temporal LS. However, what
is hidden in such a projection is that for a given number of bits,
i.e. a given branch, an infinity of different arrangement and relative
distances exists. The capacity to store information of this regenerative
memory can be understood intuitively. Since the temporal extension
of the excitable orbit is defined by its lethargic time $T_{l}$,
the maximal amount of elements that can be stored in the time delay
is the integer closest to $N\sim\tau/T_{l}=8$, in good agreement
with numerics where we found $N=7$ leading to $2^{7}=128$ possible
configurations.

\begin{figure*}
\begin{centering}
\includegraphics[clip,width=2\columnwidth]{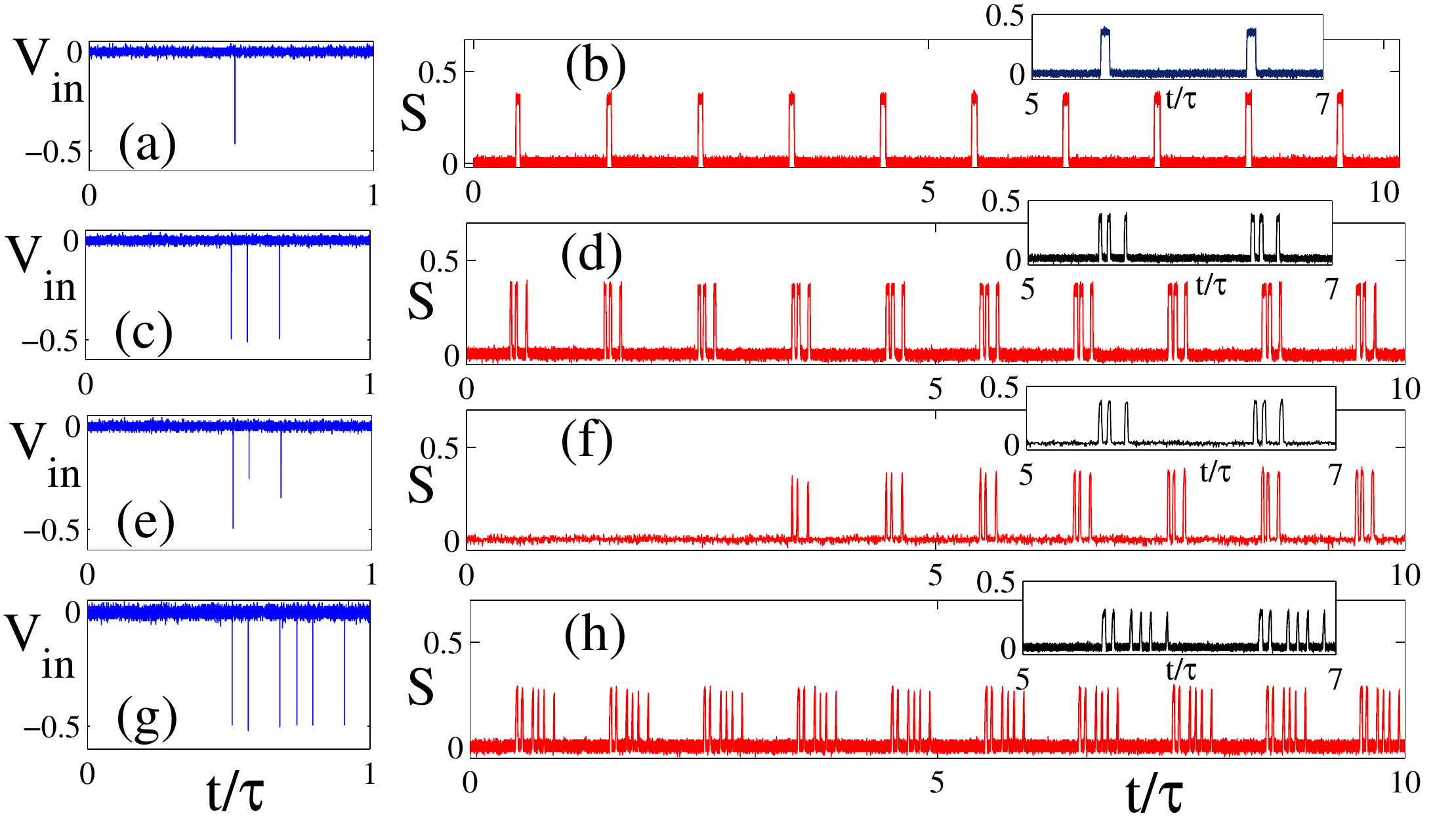}
\par\end{centering}

\caption{(Color online) Experimental time traces of showing writing and storage
of binary-coded data streams in a $\tau=100\,\mu$s cavity round-trip
time. (Left) data streams input, and (right) laser photo-detected
optical output. Sequence of (a-b) 1-bit, (c-d) 4-bit (1101), (e-f)
the same 4-bit but with a degraded input which yields the same state
in the memory. In panel f) the absence of transient is visible and
(g-h) a one byte (11011101) sequence. \label{Fig4} }
\end{figure*}

A further proof of the mutual independence of the temporal LS can
be found in a two dimensional pseudo-spatial representation depicted
in Fig.~\ref{Fig3} in which the relative motion of the various bits
of information over longer time intervals are best observed. In this
co-moving reference frame, the horizontal axis is a space-like coordinate
that allows to localize the position of the pulses within a given
round-trip while the vertical coordinate corresponds to the slow temporal
evolution of the system over many round-trips. The mutual independence
of these pulses, as demonstrated for instance by their uncorrelated
random motion in the presence of noise confirm their nature of Localized
Structures. Here, we also demonstrate the nucleation and annihilation
process in Fig.~\ref{Fig3}c) simply by including an extremely large
amount of noise. One also notice here a property of the utmost importance:
the almost total absence of transients: the localized bits of information
can be perfectly written and erased in a single round-trip, at variance
with the results of \cite{LCK-NAP-10,MJB-PRL-14} where transients
representing tens of round-trip are necessary before a stabilization
of the waveform. From an application point of view, our time-delayed
neuromorphic photonic memory presents the extraordinary advantage
to allow writing and erasing information at a rate comparable to the
nominal reading rate. As such, we believe that the possibility to
harness the unique properties of the excitable response to be of great
importance and to cross boundaries between specific fields. Finally,
the existence of the lethargic time induces an effective repulsion
between nearest bits of data when they get too close, thereby ensuring
signal integrity.

We assessed experimentally the robustness of the writing and storage
process by employing several temporal bit patterns, see methods A.
Figure \ref{Fig4}0) shows an example of complete regeneration using
a single trigger even composed of one bit (1). Besides the fact that
a single triggers is able to generate a perfectly formed train of
identical pulses one notices the almost complete absence of transient
in Fig.~\ref{Fig4}0). The regeneration of a two bits (11) and a
four-bit (1101) pattern in depicted in Fig.~\ref{Fig4}(a-b). As
mentioned previously, the bits must be separated at least by the lethargic
time $T_{l}$ of the excitable system. Moreover, it is worth noticing
that triggering an identical bit sequence can be done with an strongly
degrated initial pattern, as visible in Fig.~\ref{Fig4}c). This
demonstrate that our system performs single-pass-healing by restoring
and self-adjusting the received bits to a fixed amplitude, confirming
that the nature of the excitable response renders the regenerative
memory almost insensitive (in a certain range) to the exact shape
or the amplitude of the addressing pulses. We emphasize that the writing
and storage process is extremely robust as shown in Fig.~\ref{Fig4}
where we show the stable regeneration of a complex pattern of 8-bits
(11011101) over a time scale in the ms range.

Finally, we shed some light onto the mechanism of formation of the
multiple states of the memory by performing a bifurcation analysis
of Eqs.~(\ref{eq:FHN1},\ref{eq:FHN2}). We start by considering
the extend of the multi-stable region as a function of $\beta\in[\beta_{m}\left(\eta\right),\beta_{M}\left(\eta\right)]$
in Fig.~(\ref{Fig2}). In presence of feedback, we notice in Fig.~(\ref{Fig2})
that for $\beta<\beta^{h}\left(\eta\right)=1.018$ only the fully
developed temporal pattern with 7 equi-spaced bits is a stable solution.
We identified this lowest limit $\beta^{h}\left(\eta\right)$ as an
Andronov-Hopf bifurcation at which the uniform state, i.e. the solution
with 0 LS, becomes unstable thereby explaining why for $\beta<\beta^{h}$
the only possible states are the ones with the largest number of elements
since this background oscillation impedes the existence of empty uniform
regions. With $\eta=0.18$, only the uniform state subsists for values
of the bias $\beta>\beta_{M}\left(\eta\right)=1.3$. The upper limit
of the multi-stable region is governed by the following phenomenon.
Since the excitability threshold increases with $\beta$, it becomes
eventually too large to be overcome for the delayed replica of the
excitable orbit scaled by a factor $\eta$, suggesting that the solution
disappear as Saddle Node Bifurcations of limit cycles. 

\begin{figure*}
\includegraphics[width=2\columnwidth]{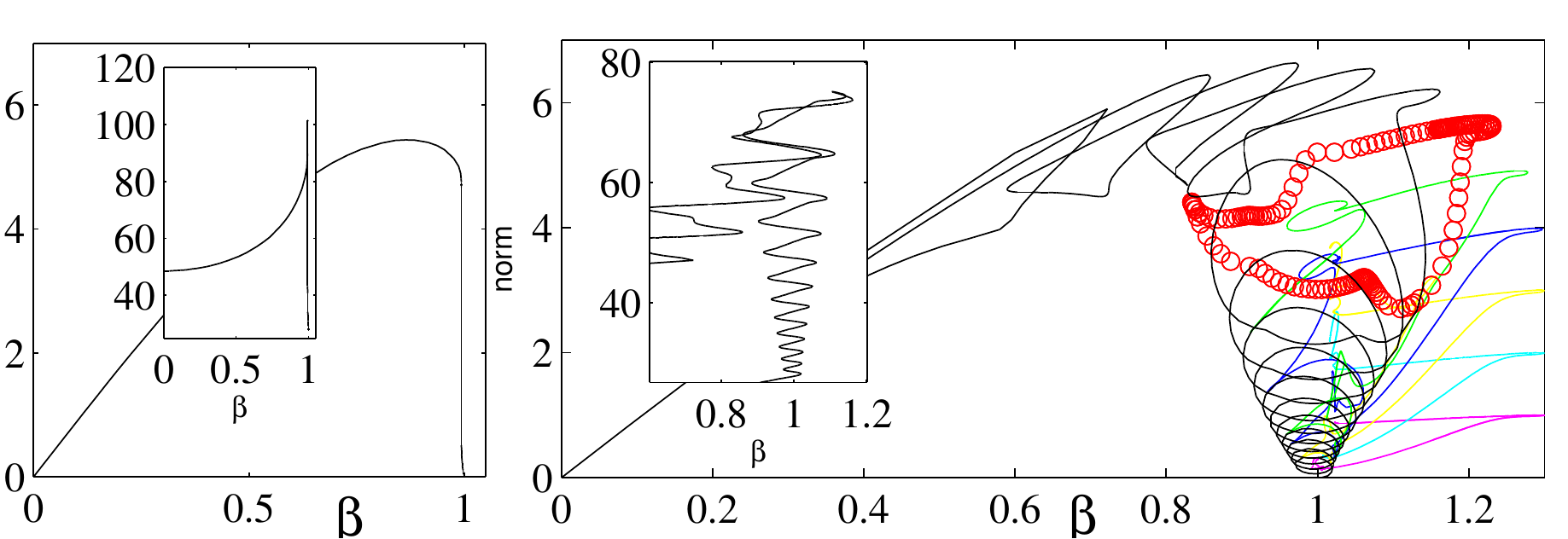} \caption{(Color online) Left: Amplitude of the periodic solutions and variation
of the period along the branches of (inset) for the FHN system without
feedback i.e. $\eta=0$. Right: Same diagram with $\eta=0.18$ the
colors correspond to the branches of solutions with 1,2,...7, equi-spaced
localized structures. \label{Fig5} }
\end{figure*}

The full bifurcation diagram of all the multi-LS solutions was obtained
with DDE-BIFTOOL \cite{DDEBT} and is depicted in Fig.~\ref{Fig5}b.
In the absence of delayed feedback (left) an Andronov-Hopf bifurcation
occurs at $\beta^{*}=1$ where weakly nonlinear oscillations rapidly
develops in large amplitude relaxation oscillations via the so called
canard phenomenon \cite{BCD-CM-81}. This explains why the bifurcation
seems to be vertical in Fig.~\ref{Fig5}a. Here, it is also important
to notice that the period is a strongly evolving function of $\beta$
(see inset in Fig.~\ref{Fig5}a). The bifurcation scenario changes
dramatically with $\eta\neq0$. The dominant periodic branch develops
a large number of folds as apparent in Fig.~\ref{Fig5}b. We stress
that this folded branch, says at $\beta=1.1$, corresponds to the
solution with a maximal number of bit within the time delay, i.e.
the trace depicted in Fig.~\ref{Fig2}d with $N=7$, sometimes called
the fully developed pattern or the maximal order solution within the
context of spatially extended systems. We have been able to identify
two different scenarios. Some branches of solution like e.g. the one
with $N=6$ in Fig.~\ref{Fig5}b) in red, can be disconnected from
the rest of the web of solution. As such they form isolae and appear
as hardly detectable saddle-node bifurcation of limit cycles. Some
other branches, like e.g. the one with $N=4$ in blue occurs as resonant
Neimark-Sacker bifurcations over the principal branch. As a function
of the bifurcation parameters $\tau$ and $\eta$, we also found that
disconnected branches would reconnect over the maximal order solution
branch and disconnect. Finally, the stability analysis of the periodic
solutions with $N$ LS exhibited $N$ quasi-degenerated Floquet multipliers
close to unity, demonstrating the independence of the various LS as
in \cite{GJT-NC-15}.

In conclusion, we have demonstrated a regenerative memory based on
the paradigm of neuronal dynamics and considering the time delay of
propagation of the signal. We disclosed the complex bifurcation diagram
from which localized bit of in formations arise from the interplay
between the excitable dynamics and the time delay. Our experimental
realization comprises a monolithic integrated resonant tunneling diode
photo-detector directly modulating a laser diode, and an optical fiber
in a delayed feedback loop providing arbitrary large information storage
capacity. The applications of the all-or-none response of the excitable
RTD to optoelectronic logic are quite direct: by adding a second,
external optical beam onto the photo-detector, one can perform an
AND operation between the bit buffer within the fiber and the external
modulated beam. This novel system constitute an ideal support for
bits in a buffer memory employing standard silica fibers combining
several functions like storage, reshaping/healing and XOR operation.
Although the present proof of concept memory operates in the MHz range,
lethargic times in the few tens of ps range can be achieved by reducing
the parasitics of the integrated circuit such as wire bonding connections
as demonstrated in \cite{RJI-OE-13,RJI-OQEL-14}. This will strongly
reduce the fiber length required to realize the optical buffer and
therefore dramatically increase the attainable bit rate, enabling
practical applications and future monolithic photonic integration.
\begin{acknowledgments}
We would like to thank Gary Ternent, University of Glasgow, UK, for
the fabrication of the RTD devices employed in this work and thank
Raquel Luis for the artistic representation of the neuron. B.R. and
R.A. acknowledge support from the CEOT and the FCT, Lisboa, Portugal.
S.B. acknowledges support from Région Provence-Alpes-Côte d'Azur through
grant number DEB 12-1538. J.J. acknowledges useful discussion with
S. Balle and financial support from the Ramon y Cajal program, project
RANGER (TEC2012-38864- C03-01) and the Direcció General de Recerca
de les Illes Balears co-funded by the European Union FEDER funds. 
\end{acknowledgments}

\subsection*{Author Contributions }

R. Avó performed the experimental characterization under the supervision
of B. Romeira and J. Figueiredo. B. Romeira and J. Figueiredo devised
the experiment. J. Javaloyes developed the theory and the numerical
analysis and wrote the manuscript together with S. Barland and B.
Romeira. All the authors participated to the interpretation of the
results.

\subsection*{Materials \& Correspondence }

Correspondence and requests for materials should be addressed to Julien
Javaloyes (email: julien.javaloyes@uib.es).

\section*{IV. Method}

\subsection{Setup details}

The experimental realization of the regenerative memory consists in
an optoelectronic feedback loop containing a c.w. laser diode, an
optical fiber delay line, and an RTD-PD. An erbium doped fiber amplifier
(EDFA) is employed in the loop to compensate the losses of coupling
and decoupling light from the RTD-PD and LD chips but can be avoided
employing RTD-PD layer structures optimized for improved photo-detection
response (estimated to be $<0.2$ A/W). The RTD-PD layer structure
currently used consists of a 10 nm wide AlAs/InGaAs/AlAs double barrier
quantum well (DBQW) structure. The epitaxial layers includes two InGaAlAs
regions surrounding the DBQW that act as light absorbing layers for
wavelengths around 1.55 $\mu$m. The LD device was a InGaAsP multi-quantum-well
active region with an InP:Fe doped uncoated buried heterostructure
and ridge mesas, with centre wavelength emission at $\sim$$1.55\,\mu$m,
threshold current $\sim6$ mA \cite{RJI-OE-13}, and 3-dB modulation
bandwidth in excess of 10 GHz.

In this work, for purposes of demonstration and experimental convenience,
the data signals were injected electrically using a bias-T. The binary-coded
data streams were generated using an arbitrary function generator
(Tektronix AFG3251C). The signals could be also injected optically,
taking advantage of the optical input port of the RTD-PD, and therefore
buffering can be achieved using either electrical or optical incoming
data. By operating the system in the excitable regime close to the
NDR region, one would be able to store and regenerate optical bits
of information in the fibre, the empty region signaling the \textquotedbl{}0\textquotedbl{}
bits and the excitable optoelectronic pulses the \textquotedbl{}1\textquotedbl{}.
When these bits are re-injected into the photo-diode they trigger
the generation of a new excitable cycle. This regenerative mechanism
occurs after each round trip in the fibre which is extremely robust
since even if the bit sequence is strongly deteriorated the all-or-none
response of the excitable RTD allows for a perfect regeneration. We
employed a very long optical fiber loop providing $46\,\mu$s cavity
round-trip time, $\tau$. The time $\tau$ was chosen in order that
the memory buffer is much larger than the typical excitable lethargic
time, $T_{l}$, of the RTD-PD-LD excitable system, where in this case
the measured $T_{l}$ is around 500 ns.

\subsection{Theoretical model}

We have analyzed our experimental results in the framework of delayed
feedback nonlinear dynamical model systems employing a Liénard equation
describing the RTD-PD excitable oscillator \cite{RJF-JQE-13,RJI-OQEL-14}
coupled to the single mode rate equations modeling the laser intensity
and its population inversion. Regeneration is induced in such dynamical
system by a delayed feedback term into the current source of the RTD-PD
proportional to the intensity of the laser output. The Liénard oscillator-laser
diode dynamical system is given by the following dimensionless coupled
delay differential equations (DDEs):

\begin{eqnarray}
\mu\dot{v} & = & i-f(v)-\eta s(t-\tau_{d}),\label{eq:voltage}\\
\mu^{-1}\dot{i} & = & \upsilon_{0}-\gamma i-v,\label{eq:current}\\
\tau_{n}^{-1}\dot{n} & = & \frac{i}{i_{th}}-n-\frac{n-\delta}{1-\delta}\left(1-\epsilon s\right)s,\label{eq:carrier}\\
\tau_{p}^{-1}\dot{s} & = & \frac{n-\delta}{1-\delta}\left(1-\epsilon s\right)s-s.\label{eq:photon}
\end{eqnarray}

Equations (\ref{eq:voltage})-(\ref{eq:photon}) represent the system
of equations of the RTD-PD-LD with optical feedback control through
the variable $s(t-\tau_{d})$ where $\tau_{d}$ is the time-delay
with respect to the dimensionless time $t$; time is normalized to
the characteristic $LC$ resonant tank frequency, $\omega_{0}=(\sqrt{LC})^{-1}$,
hence $\tau=\omega_{0}t$. The feedback strength $\eta$ parameter
depends on RTD-PD detection characteristics and the fraction of the
laser optical output power reinjected into the delayed feedback loop.
Equations (\ref{eq:voltage})-(\ref{eq:photon}) represent the Liénard
oscillator where $v$ and $i$ are the dimensionless voltage and current
variables, respectively. The function $f(v)$ describes the nonlinear
I-V curve, $R=\gamma(I_{0}/V_{0})$, and $\mu=V_{0}/I_{0}\sqrt{C/L}$
is a dimensionless parameter.

Equations (\ref{eq:voltage})-(\ref{eq:photon}) are the dimensionless
rate equations describing LD normalized photon $s(t)$ and injected
carrier $n(t)$ densities. The charge carrier $n(t)$ in Eq.~(\ref{eq:photon})
is normalized to threshold providing that $\delta=N_{0}/N_{th}$,
where $N_{0}$ is the carrier density for transparency, and $N_{th}$
is the threshold carrier density; $\epsilon$ stands for the dimensionless
laser gain saturation. The dimensionless laser diode threshold current
is $i_{th}$; The parameters $\tau_{n}$ and $\tau_{p}$ come from
the time rescaling.

The prediction of such a system very good agreement with the experimental
results allowing to predict a theoretical limit of operation of several
Gb/s for such buffer configuration. In the limit case where the pulses
are broad and the dynamics is slow as compared to the relaxation oscillation
frequency of the laser, it is possible to adiabatically eliminate
the equations for $\left(s,n\right)$. As such the delayed intensity
$s\left(t-\tau\right)$ becomes proportional to the bias of the laser
device $i\left(t\right)$. As a last step, the nonlinear function
$f\left(v\right)$ has to be expanded in Taylor series around the
center of the NDR. Neglecting the second order term yielding the asymmetry
of $f\left(\upsilon\right)$ in the expansion and cutting the expansion
to third yields the FHN equations presented in the main manuscript,
after further trivial rescaling.


\begin{thebibliography}{10}

\bibitem{V-BR-72}
Hendrik Van~Der Loos and Edmund~M. Glaser.
\newblock Autapses in neocortex cerebri: synapses between a pyramidal cell's
  axon and its own dendrites.
\newblock {\em Brain Research}, 48(0):355 -- 360, 1972.

\bibitem{HK-IJBC-04}
Christoph~S. Herrmann and Andreas Klaus.
\newblock Autapse turns neuron into oscillator.
\newblock {\em International Journal of Bifurcation and Chaos},
  14(02):623--633, 2004.

\bibitem{F-NRN-09}
Monica~Hoyos Flight.
\newblock Neuromodulation: Exerting self-control for persistence.
\newblock {\em Nat Rev Neurosci}, 10(5):316--316, May 2009.

\bibitem{B-NRN-09}
Tiago Branco and Kevin Staras.
\newblock The probability of neurotransmitter release: variability and feedback
  control at single synapses.
\newblock {\em Nat Rev Neurosci}, 10(5):373--383, May 2009.

\bibitem{JCE-NL-10}
Sung~Hyun Jo, Ting Chang, Idongesit Ebong, Bhavitavya~B. Bhadviya, Pinaki
  Mazumder, and Wei Lu.
\newblock {Nanoscale Memristor Device as Synapse in Neuromorphic Systems}.
\newblock {\em {Nano Letters}}, {10}({4}):{1297--1301}, {Apr} {2010}.

\bibitem{ILB-FRNE-11}
Giacomo Indiveri, Bernabe Linares-Barranco, Tara~Julia Hamilton, Andre van
  Schaik, Ralph Etienne-Cummings, Tobi Delbruck, Shih-Chii Liu, Piotr Dudek,
  Philipp Hafliger, Sylvie Renaud, Johannes Schemmel, Gert Cauwenberghs, John
  Arthur, Kai Hynna, Fopefolu Folowosele, Sylvain Saighi, Teresa
  Serrano-Gotarredona, Jayawan Wijekoon, Yingxue Wang, and Kwabena Boahen.
\newblock {Neuromorphic silicon neuron circuits}.
\newblock {\em {Frontiers in neuroscience}}, {5}, {2011}.

\bibitem{MAA-SCI-14}
Paul~A. Merolla, John~V. Arthur, Rodrigo Alvarez-Icaza, Andrew~S. Cassidy, Jun
  Sawada, Filipp Akopyan, Bryan~L. Jackson, Nabil Imam, Chen Guo, Yutaka
  Nakamura, Bernard Brezzo, Ivan Vo, Steven~K. Esser, Rathinakumar Appuswamy,
  Brian Taba, Arnon Amir, Myron~D. Flickner, William~P. Risk, Rajit Manohar,
  and Dharmendra~S. Modha.
\newblock {A million spiking-neuron integrated circuit with a scalable
  communication network and interface}.
\newblock {\em {Science}}, {345}({6197}):{668--673}, {Aug 8} {2014}.

\bibitem{TNS-JLT-14}
A.N. Tait, M.A. Nahmias, B.J. Shastri, and P.R. Prucnal.
\newblock Broadcast and weight: An integrated network for scalable photonic
  spike processing.
\newblock {\em Lightwave Technology, Journal of}, 32(21):4029--4041, Nov 2014.

\bibitem{KJL-NL-11}
Duygu Kuzum, Rakesh G.~D. Jeyasingh, Byoungil Lee, and H.~S~Philip Wong.
\newblock Nanoelectronic programmable synapses based on phase change materials
  for brain-inspired computing.
\newblock {\em Nano Letters}, 12(5):2179--2186, Jun 2011.

\bibitem{E-RPP-98}
Bard Ermentrout.
\newblock Neural networks as spatio-temporal pattern-forming systems.
\newblock {\em Reports on Progress in Physics}, 61(4):353, 1998.

\bibitem{BT-PRE-03}
Nikola Buri{\'c} and Dragana Todorovi{\'c}.
\newblock Dynamics of fitzhugh-nagumo excitable systems with delayed coupling.
\newblock {\em Phys. Rev. E}, 67:066222, Jun 2003.

\bibitem{S-PTA-09}
Gabor Stepan.
\newblock Delay effects in brain dynamics.
\newblock {\em Philosophical Transactions of the Royal Society A: Mathematical,
  Physical and Engineering Sciences}, 367(1891):1059--1062, 2009.

\bibitem{YMG-PRE-02}
Alejandro~M. Yacomotti, Gabriel~B. Mindlin, Massimo Giudici, Salvador Balle,
  Stephane Barland, and Jorge Tredicce.
\newblock Coupled optical excitable cells.
\newblock {\em Phys. Rev. E}, 66:036227, Sep 2002.

\bibitem{KBS-PRE-10}
B.~Kelleher, C.~Bonatto, P.~Skoda, S.~P. Hegarty, and G.~Huyet.
\newblock Excitation regeneration in delay-coupled oscillators.
\newblock {\em Physical Review E}, 81(3):036204, 2010.

\bibitem{WEK-PRE-14}
Lionel Weicker, Thomas Erneux, Lars Keuninckx, and Jan Danckaert.
\newblock Analytical and experimental study of two delay-coupled excitable
  units.
\newblock {\em Phys. Rev. E}, 89:012908, Jan 2014.

\bibitem{GP-PRL-96}
G.~Giacomelli and A.~Politi.
\newblock Relationship between delayed and spatially extended dynamical
  systems.
\newblock {\em Phys. Rev. Lett.}, 76:2686--2689, Apr 1996.

\bibitem{CoulletLSinfo}
P.~Coullet, C.~Riera, and C.~Tresser.
\newblock A new approach to data storage using localized structures.
\newblock {\em Chaos}, 14:193--201, Mar 2004.

\bibitem{turing52}
A.~M. Turing.
\newblock The chemical basis of morphogenesis.
\newblock {\em Philos. Trans. R. Soc London}, 237:37, 1952.

\bibitem{NP-SelfOrg-77}
G.~Nicolis and I.~Prigogine.
\newblock {\em {Self-Organization} in Nonequilibrium Systems: From Dissipative
  Structures to Order through Fluctuations}.
\newblock Wiley, 1977.

\bibitem{DC-LNP-11}
O.~Descalzi, M.~Clerc, S.~Residori, and G.~Assanto.
\newblock {\em Localized States in Physics: Solitons and Patterns}, volume 751
  of {\em Lecture Notes in Physics}.
\newblock Springer Berlin Heidelberg, 2011.

\bibitem{sands}
Paul~B. Umbanhowar, Francisco Melo, and Swinney~Harry L.
\newblock Localized excitations in a vertically vibrated granular layer.
\newblock {\em Nature}, (382):793 -- 796, 1996.

\bibitem{Astrov2001349}
Yuri~A. Astrov and Hans-Georg Purwins.
\newblock Plasma spots in a gas discharge system: birth, scattering and
  formation of molecules.
\newblock {\em Physics Letters A}, 283(5–6):349 -- 354, 2001.

\bibitem{NAD-PSS-92}
F.-J. Niedernostheide, M.~Arps, R.~Dohmen, H.~Willebrand, and H.-G. Purwins.
\newblock Spatial and spatio-temporal patterns in pnpn semiconductor devices.
\newblock {\em physica status solidi (b)}, 172(1):249 -- 266, 1992.

\bibitem{reactiondiffusion}
Kyoung-Jin Lee, William~D. McCormick, John Pearson, and Harry~L. Swinney.
\newblock Experimental observation of self-replicating spots in a
  reaction-diffusion system.
\newblock {\em Nature}, 369:215--218, 1994.

\bibitem{PhysRevLett.52.1421}
Junru Wu, Robert Keolian, and Isadore Rudnick.
\newblock Observation of a nonpropagating hydrodynamic soliton.
\newblock {\em Phys. Rev. Lett.}, 52:1421--1424, Apr 1984.

\bibitem{PhysRevA.35.2757}
Elisha Moses, Jay Fineberg, and Victor Steinberg.
\newblock Multistability and confined traveling-wave patterns in a convecting
  binary mixture.
\newblock {\em Phys. Rev. A}, 35:2757--2760, Mar 1987.

\bibitem{BTB-NAT-02}
S.~Barland, J.~R. Tredicce, M.~Brambilla, L.~A. Lugiato, S.~Balle, M.~Giudici,
  T.~Maggipinto, L.~Spinelli, G.~Tissoni, T.~Knodl, M.~Miller, and R.~Jager.
\newblock Cavity solitons as pixels in semiconductor microcavities.
\newblock {\em Nature}, 419(6908):699--702, Oct 2002.

\bibitem{HPC-JSTQE-06}
X.~Hachair, F.~Pedaci, E.~Caboche, S.~Barland, M.~Giudici, J.~R. Tredicce,
  F.~Prati, G.~Tissoni, R.~Kheradmand, L.A. Lugiato, I.~Protsenko, and
  M.~Brambilla.
\newblock Cavity solitons in a driven vcsel above threshold.
\newblock {\em Selected Topics in Quantum Electronics, IEEE Journal of},
  12(3):339--351, 2006.

\bibitem{LCK-NAP-10}
F.~Leo, S.~Coen, P.~Kockaert, S.P. Gorza, P.~Emplit, and M.~Haelterman.
\newblock Temporal cavity solitons in one-dimensional kerr media as bits in an
  all-optical buffer.
\newblock {\em Nat Photon}, 4(7):471--476, Jul 2010.

\bibitem{GJT-NC-15}
B.~Garbin, J.~Javaloyes, G.~Tissoni, and S.~Barland.
\newblock Topological solitons as addressable phase bits in a driven laser.
\newblock {\em Nat. Com.}, 6, 2015.

\bibitem{BKY-OL-11}
Sylvain Barbay, Robert Kuszelewicz, and Alejandro~M. Yacomotti.
\newblock Excitability in a semiconductor laser with saturable absorber.
\newblock {\em Optics letters}, 36(23):4476--4478, 2011.

\bibitem{SBB-PRL-14}
F.~Selmi, R.~Braive, G.~Beaudoin, I.~Sagnes, R.~Kuszelewicz, and S.~Barbay.
\newblock Relative refractory period in an excitable semiconductor laser.
\newblock {\em Phys. Rev. Lett.}, 112:183902, May 2014.

\bibitem{MGB-PRL-14}
Francesco Marino, Giovanni Giacomelli, and Stephane Barland.
\newblock Front pinning and localized states analogues in long-delayed bistable
  systems.
\newblock {\em Phys. Rev. Lett.}, 112:103901, Mar 2014.

\bibitem{SNJ-JAP-11}
AS~Samardak, A~Nogaret, NB~Janson, A~Balanov, I~Farrer, and DA~Ritchie.
\newblock Spiking computation and stochastic amplification in a neuron-like
  semiconductor microstructure.
\newblock {\em Journal of Applied Physics}, 109, 2011.

\bibitem{RJI-OE-13}
Bruno Romeira, Julien Javaloyes, Charles~N. Ironside, Jos\'{e} M.~L.
  Figueiredo, Salvador Balle, and Oreste Piro.
\newblock Excitability and optical pulse generation in semiconductor lasers
  driven by resonant tunneling diode photo-detectors.
\newblock {\em Opt. Express}, 21(18):20931--20940, Sep 2013.
\newblock 1.

\bibitem{HH-JOP-52}
A.~L. Hodgkin and A.~F. Huxley.
\newblock A quantitative description of membrane current and its application to
  conduction and excitation in nerve.
\newblock {\em Journal of Physiology}, 117(4):500--544, 1952.

\bibitem{F-BMB-55}
Richard FitzHugh.
\newblock Mathematical models of threshold phenomena in the nerve membrane.
\newblock {\em The bulletin of mathematical biophysics}, 17(4):257--278, 1955.

\bibitem{NAY-PIRE-62}
J.~Nagumo, S.~Arimoto, and S.~Yoshizawa.
\newblock An active pulse transmission line simulating nerve axon.
\newblock {\em Proceedings of the IRE}, 50(10):2061--2070, Oct 1962.

\bibitem{L-RGE-28}
Alfred-Marie Liénard.
\newblock Etude des oscillations entretenues.
\newblock {\em Revue générale de l'électricité}, 23(1):901–912 and
  946–954, 1928.

\bibitem{RJF-JQE-13}
B.~Romeira, J.~Javaloyes, J.M.L. Figueiredo, C.N. Ironside, H.I. Cantu, and
  A.E. Kelly.
\newblock Delayed feedback dynamics of lienard-type resonant
  tunneling-photo-detector optoelectronic oscillators.
\newblock {\em Quantum Electronics, IEEE Journal of}, 49(1):31 --42, jan. 2013.
\newblock 4.

\bibitem{RJI-OQEL-14}
Bruno Romeira, Ricardo Av{\'o}, Julien Javaloyes, Salvador Balle, Charles~N.
  Ironside, and J.~M.~L. Figueiredo.
\newblock Stochastic induced dynamics in neuromorphic optoelectronic
  oscillators.
\newblock {\em Optical and Quantum Electronics}, pages 1--6, 2014.
\newblock 0.

\bibitem{HH2-JOP-52}
A.~L. Hodgkin, A.~F. Huxley, and B.~Katz.
\newblock Measurement of current-voltage relations in the membrane of the giant
  axon of loligo.
\newblock {\em The Journal of physiology}, 116(4):424, 1952.

\bibitem{P-PLA-92}
K.~Pyragas.
\newblock Continuous control of chaos by self-controlling feedback.
\newblock {\em Physics Letters A}, 170(6):421 -- 428, 1992.

\bibitem{BCD-CM-81}
Eric Beno{\^i}t, Jean~Louis Callot, Francine Diener, and Marc Diener.
\newblock Chasse au canard (première partie).
\newblock {\em Collectanea Mathematica}, 32(1), 1981.

\bibitem{keener2008}
J.~Keener and J.~Sneyd.
\newblock {\em Mathematical Physiology: I: Cellular Physiology}, volume~1.
\newblock Springer, 2008.

\bibitem{Meron19921}
Ehud Meron.
\newblock Pattern formation in excitable media.
\newblock {\em Physics Reports}, 218(1):1 -- 66, 1992.

\bibitem{Tyson1988327}
John~J. Tyson and James~P. Keener.
\newblock Singular perturbation theory of traveling waves in excitable media (a
  review).
\newblock {\em Physica D: Nonlinear Phenomena}, 32(3):327 -- 361, 1988.

\bibitem{MJB-PRL-14}
M.~Marconi, J.~Javaloyes, S.~Balle, and M.~Giudici.
\newblock How lasing localized structures evolve out of passive mode locking.
\newblock {\em Phys. Rev. Lett.}, 112:223901, Jun 2014.

\bibitem{DDEBT}
Koen Engelborghs, Tatyana Luzyanina, and Giovanni Samaey.
\newblock Dde-biftool v. 2.00: a matlab package for bifurcation analysis of
  delay differential equations.
\newblock Technical report, Department of Computer Science, K.U.Leuven,
  Belgium., 2001.

\end{thebibliography}

\def\url#1{}

\end{document}